\documentstyle[12pt,epsfig]{article}
\def \Pom {{\hspace{ -0.1em}I\hspace{-0.25em}P}}
\begin{document}

\begin{flushright}
ITP-96-13E
\end{flushright}

\vskip 2.cm
\centerline{\bf REGGE POLE MODEL FOR VECTOR MESON}
\centerline{\bf PHOTOPRODUCTION AT
$HERA$}

\vskip 0.4cm
\centerline{L.L. Jenkovszky, E.S.Martynov and F.Paccanoni}
\vskip 0.3cm

\begin{center}
{\it Bogoliubov Institute for Theoretical Physics,\\
Kiev 143, Ukraine}\\
E-mail: jenk@gluk.apc.org, \, martynov@gluk.apc.org, \, paccanoni@padova.infn.it
\end{center}
\vskip 15.0pt

\begin{abstract}
Recent $HERA$ data on the photoproduction of vector mesons are analysed
within a "soft", dipole pomeron model. We argued
that the data on $\sigma^{\gamma\rightarrow J/\psi}_{el}$ is consistent
with a soft pomeron, the apparent rapid increase resulting from the
non-asymptotic effects due to the delayed asymptotics of $\sigma_{el}$
with respect to $\sigma_{tot}$.
\end{abstract}

\newpage
\vskip 2.truecm
\section{}
~

Studies of exclusive -- diffractive and nondiffractive --
photoproduction of vector mesons at $HERA$ have many important
new aspects, among which is the applicability of the Regge
pole theory for virtual particles, the main subject of the present
paper. We construct an explicit Regge-behaved model for virtual
particles and discuss its observable consequences in the
light of the relevant $HERA$ measurements.

Let us remind that the basic diagram in question is that shown in
Fig. 1, where $s$ and $t$ are the Mandelstam variables and $Q^2=-q^2$
is the virtuality of the external particle. In most of the cases
studied previously, only two variables were present: 1. $s$ and $t$
(with $Q^2=m^2$) for (on-shell) hadronic (exclusive) reactions,
and 2.  $s$ and $Q^2$ (with $t=0$) for forward virtual Compton
scattering, related by unitarity to deep inelastic (inclusive)
scattering.
In exclusive virtual photoproduction, on the other hand one has the
unique situation when all three variables, $s,\ t$ and  $Q^2$, meet.
Moreover, in the photoproduction of heavy mesons, e.g. $J/\psi$,
their masses become another important parameter. In that case, one
introduces the sum
$$\tilde q^2=(Q^2+M^2) \eqno(1)$$
as a new  ``scaling'' variable, i.e. one assumes that the large
mass of the external particle plays the same role as the (photon)
virtuality, and consequently $J/\psi$ production in this sense is
``hard''.
\vskip 1.cm
\parbox{5.5cm}{
\vskip 3.8cm
\begin{center}
\epsfig{file=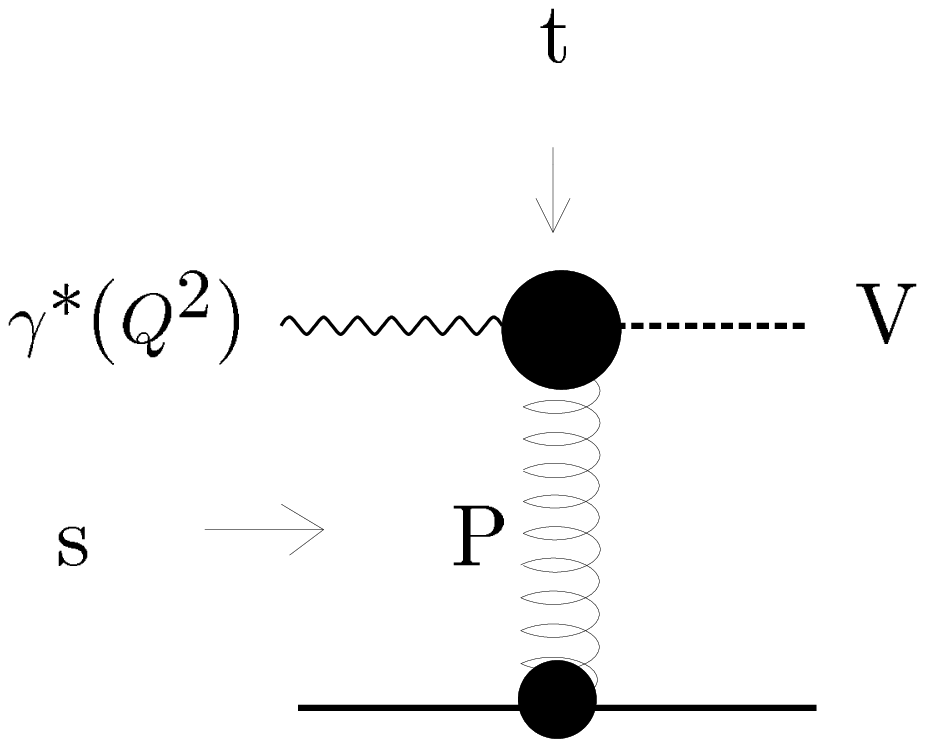,width=5.cm}
\end{center}
\vskip .5cm
Fig.1. Photoproduction of vector meson.}\ \quad \
\parbox{9.5cm}{
\begin{center}
\epsfig{file=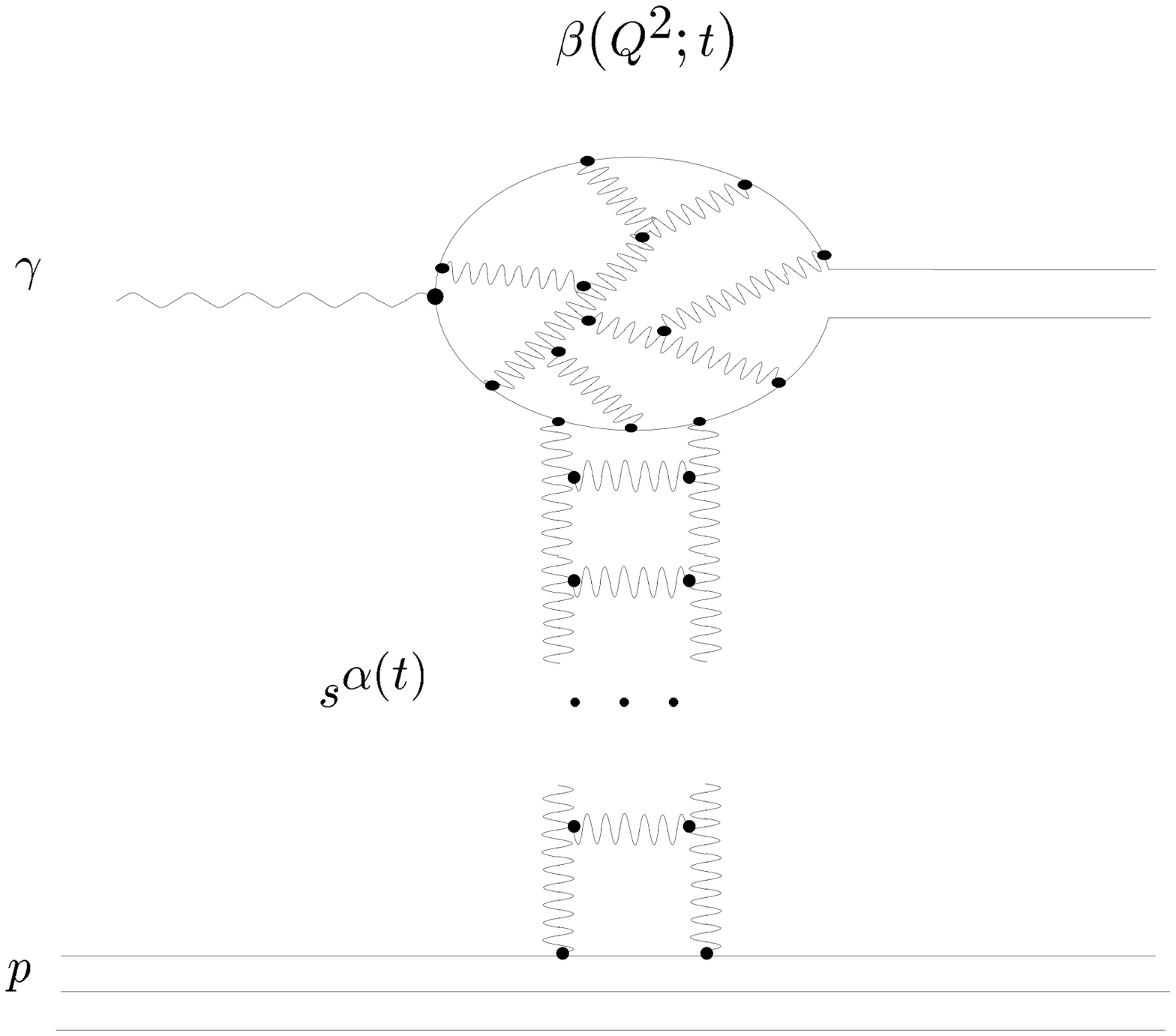,width=9.0cm}
\end{center}
\vskip .5cm
Fig.2. Quark-gluon picture of vector meson photo\-pro\-duc\-tion.
Factorization implies that $Q^2$ evolution comes from the upper vertex}
\vskip 1.cm

Let us now discuss the "hardness" of the pomeron. According to the
$HERA$ data,
$$\sigma_{el}^{\gamma V}\sim W^n, \eqno (2)$$
where $n=n(\tilde q^2)$ tends to increase with increasing
``virtuality'' $\tilde q^2$.  Typically, $n\simeq 0.22 \div 0.32$ for
elastic photoprocution of light vector mesons and $n\simeq 0.8$ for
deep inelastic photoproduction and/or production of $J/\psi$. This
phenomenon was usually interpreted as the manifestation of two different
pomerons --
a ``soft'' $(n=4\delta\simeq0.3)$ and a ``hard'' one
$(n\simeq 0.8)$, where $\delta=\alpha(0)-1$ is related to the
pomeron intercept.

In this paper we present a different point of view. In our opinion
there is only one pomeron in Nature, the variation
with $\tilde q$ coming from the relevant vertex as shown in
Fig. 2.

In this approach, the pomeron
propagator is essentially non-perturbative, while the $\bar
q$-dependence is given either by the QCD evolution or by
the relevant phenomenological vertex function.
Factorization is an important property of the theory. It is valied in
the case of exchange of a simple pole. In our model factorization holds
for the partial amplitude (in the $(j,t)-$representation but not in the
$(s,t)-$one).

Another important aspect of the new $HERA$ data is the possibility to
detect a clear pomeron signal in the photoproduction of $\phi$
and $J/\psi$, where - by the Okubo-Iizuki-Zweig rule - the
contribution from secondary trajectories is suppressed.
\medskip

\section{}
~

As it is well known, the Regge pole model was formulated within the
S-matrix theory, applicable only on the mass shell.
Attempts to generalize the Regge pole model for currents
were undertaken long ago, with little success, however. A discussion of this problem can be
found e.g. in the book \cite{Collins}. Here we take a pragmatic point
attitude in constructing an optimal model combining the known
principles of the theory with the observations.

A partial (for $t=0$) solution to this problem was suggested in the
paper \cite{JMP}, in which the inclusive deep inelastic data in a
wide range of the variable were fitted by a Regge-type model with a
phenomenological expression for the $Q^2$-dependent vertex function.
The next step is to extend this model to include $t$-dependence as well.

A related approach was pursued in refs. \cite{Cap,Haak},
where the $Q^2$ dependence was introduced in the parameters of
the Regge pole, paticularly in the pomeron intercept $\alpha_0(Q^2)$.
Arguments were presented in that papers on how unitarity
constrains this dependence. Notably, $\alpha_0$ rises with $Q^2$ such
as to meet the effect observed at $HERA$.

Contrary to the above \cite{Cap} and related papers \cite{DL,Buch,Amun}
based on the so-called supercritical pomeron, we use a dipole pomeron
pole with unit interecept. Logarithmically rising total cross sections
and small-$x$ structure functions are typical of this class of models.
They do not violate the Froissart bound and therefore need not to be
unitarized. $Q^2$ dependence will be introduced in the parameters of
the residue and in the scaling parameter $s_0$, but not in pomeron
intercept.

Most generally the dipole pomeron model generalized for virtual
external particles can be written as follows \cite{JMP}
$$A(W,t;\tilde q^2) = \Pom (W,t;\tilde q^2) + f(W,t;\tilde q^2) + ...
,$$ where $\Pom$ is a Pomeron contribution
$$\Pom (W,t;\tilde q^2) =
ig_{0}(t;\tilde q^2) \Bigl(\frac{-is}{s_{0}(\tilde
q^2)}\Bigr)^{\alpha_{\Pom}(t)-1} + ig_{1}(t;\tilde
q^2)ln(\frac{-is}{s_{1}(\tilde q^2)}) \Bigl(\frac{-is}{s_{1}(\tilde
q^2)}\Bigr)^{\alpha_{\Pom}(t)-1}$$
with $s=W^{2}$ and
$$g_{i}(t;\tilde q^2) =
g_{i}(\tilde q^2)\exp (b_{i}(\tilde q^2)t).$$
A contribution of
the $f$-reggeons is written similarly:
$$f(W,t;\tilde q^2) =
ig_{f}(t;\tilde q^2) \Bigl(\frac{-is}{s_{f}(\tilde
q^2)}\Bigr)^{\alpha_{f}(t)-1}.$$
It is important to note that in this model the intercept of the
trajectory equals to 1
$$\alpha_{\Pom}(0) = 1.$$
Thus the model does
not violate the Froissart-Martin bound.

The parameters of $\alpha_{\Pom}(t),\alpha_{f}(t)$ are universal,
independent of the reaction, while $g_{i},b_{i},s_{i}$ are
reaction-dependent. In hadronic phenomenology $g_{i}$ and
$b_{i}$ are constants.  Here they should be replaced by
$Q^{2}$-dependent functions.
The slope $B(s,t;Q^{2})=2b_{i}(Q^{2})+2\alpha^{'}_{\Pom}ln(s/s_{i}(Q^2)$
contains a
universal energy-dependent term, while the parameter $b_{i}(Q^{2})$
is responsible for the quark content (quark number, masses and
flavors).

In this paper we concentrate on the dynamics of the reactions rather
than the quark and symmetry relations between them. We just note
that the local slope of diffractive $J/\psi$ production is
much smaller than that for other mesons since the heavier
$J/\psi$ is much more compact than the rest of the mesons are.

\medskip

\section{}
~

For $\rho,\ \omega$ and $\phi$ photoproduction we write the
scattering amplitude as a sum of a pomeron and $f$ contribution. In
the case of $J/\psi$, non-pomeron contributions are suppressed due to
the Okubo-Iizuki-Zweig rule. Notice, that although $\phi$ also consists
of strange quarks, it receives - albeit a small - contribution from
secondary reggeons due to the $\omega-\phi$-mixing.
By setting $A="\Pom"+"f"+...,$ we get the integrated elastic cross
section $\sigma_{el}$:
$$\sigma^{\gamma^{*}p\rightarrow
Vp}_{el} = 4\pi\int\limits_{-\infty}^{0}dt|A^{\gamma^{*}p\rightarrow
Vp}(W,t;Q^{2})|^{2}$$
$$ = 4\pi \biggl \{ \frac{g_{0}^{2}}{2B_{0}} +
\frac{g_{0}g_{1}}{B_{0}+B_{1}}\xi +
\frac{g_{1}^{2}}{2B_{1}}(\xi^{2}+\pi^{2}/4) +
\frac{g_{f}^{2}}{2B_{f}}
\biggl (\frac{s}{s_{f}}\biggr )^{2\alpha_{f}(0)-2}$$
$$ + 2g_{f}\biggl (\frac{s}{s_{f}}\biggr )^{\alpha_{f}(0)-1}
\biggl [g_{1}\frac{C_{f} (\xi (B_{1}+B_{f})+D\pi /2)
 - S_{f}(\xi D-(B_{1}+B_{f})\pi /2)}{
(B_{1}+B_{f})^{2}+D^{2}}$$
$$ + g_{0}\frac{C_{f}(B_{0}+B_{f})-S_{f}D}{(B_{0}+B_{f})^{2}+D^{2}}
\biggr ]\biggr \},\eqno(3)$$
where
$$C_{f} = \cos(\frac{\pi}{2}(\alpha_{f}(0)-1)),\qquad
S_{f} = \sin(\frac{\pi}{2}(\alpha_{f}(0)-1))$$
$$B_{i} = \alpha^{'}ln (s/s_{i}) + b_{i}, \qquad i= 0,1,f,$$
$$D = \frac{\pi}{2}(\alpha^{'}_{P}-\alpha^{'}_{f}),$$
$$\xi = ln(s/s_{1}).$$
Generally, $g_{i}$ and $b_{i}$ may depend on
$\tilde q^{2}$. However here we consider only the case $Q^2=0$, therefore
we put $g_i,\,b_i = const$ and $s_i=1GeV^2$.

Let us first discuss the case
of $J/\psi$. Familiar extrapolations from the old, low-energy data to
those from $HERA$ $\sigma_{el}\sim W^{0.8}$ give a rate of increase
much larger than that for total photoproduction, which is \cite{H1,ZEUS}
$\sigma_{tot}\sim W^{0.2}$. Since the partial cross section for
$J/\psi$ production makes only part of the total cross section the
continuation of this trend sooner or later will violate unitarity.
Hence one has to assume that the cross sections have not yet reached
their asymptotic regimes.  This means that either the total cross
section will rise faster or that the expected asymptotic rise for
$J/\psi$ is slower than that quoted above.

To illustrate our arguments, we present a fit (without any
minimization procedure) to the data on $\rho, \omega, \phi$ and
$J/\psi$ photoproduction, as well as $\gamma p$ total cross-section,
based on the present model (3).  Notice that the sharp rise in the case
of $J/\psi$ at low $\sqrt{s}\leq 20 GeV$ is a preasymptotic effect and
it is not indicative of the "hardness" of the pomeron. Besides this, the
threshold at $s=(m_p^2+m_{V}^2)$ must be taken into account in order
to describe the experimental data for $J/\phi$  and $J/\phi$
photoproduction.  For this case we multiply the amplitude (3) by factor
$[(1-(m_p^2+m_{V}^2)/s]^{\nu_{V}}.$  Comparison with the data is
illustrated in Fig.3, with the parameters given in the Table.

\begin{center}
\vskip 1.cm
\hskip -.5cm
\epsfig{file=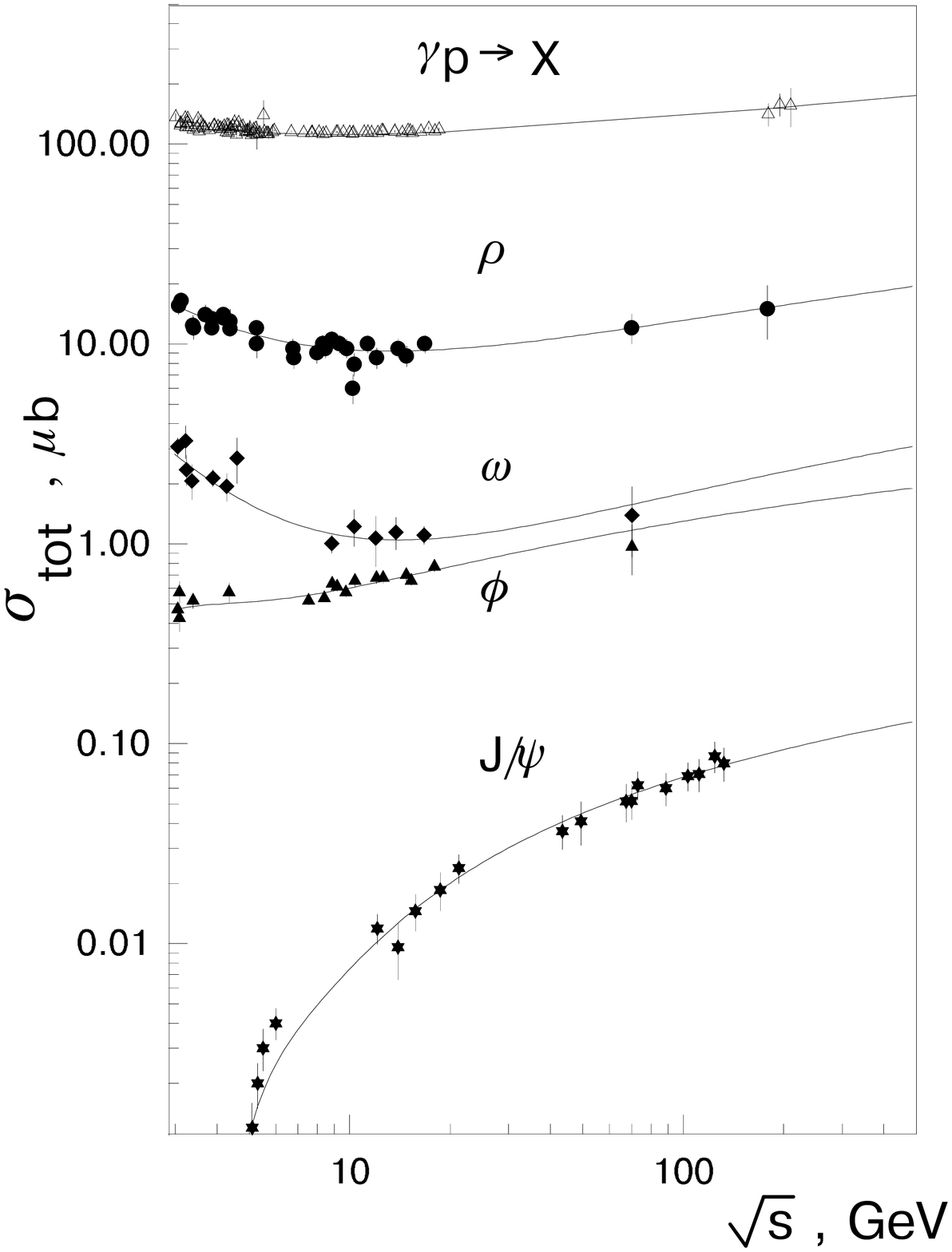,height=10.cm}

Fig.3. Elastic vector meson production in the model (3).
\end{center}
\vskip 0.5 cm
Table. Parameters used in Exp.(3) for the description of the experimental data.
The values of the slopes $\alpha'_{\Pom}=0.25GeV^{-2},\,
\alpha'_{f}=0.85GeV^{-2}$ and the intercept $\alpha_{f}(0)=0.8$ are
choosen from hadronic phenomenology.
\vskip1.cm
\begin{tabular}{|c|c|c|c|c|c|c|}
\hline
& & & & & & \\
&  $g_{0}, \mu b$ & $g_{1}, \mu b$ & $g_{f}, \mu b$ & $b_{0}, GeV^{-2}$ &
$b_{1}, GeV^{-2}$ &  $b_{f}, GeV^{-2}$\\
& & & & & & \\
\hline
$\gamma p\rightarrow X$ & -2.18 & 0.70 & 9.23 & - & - & - \\
& & & & & & \\
\hline
$\gamma p\rightarrow \rho p$ & -30.10 & 7.75 & 118.4 & 3.89 & 1.83 & 2.32 \\
& & & & & & \\
\hline
$\gamma p\rightarrow \omega p$ & -24.88 & 3.84 & 62.69 & 3.03 & 1.48 & 2.50 \\
& & & & & & \\
\hline
$\gamma p\rightarrow \phi p$ & -5.82 & 1.84 & - & 0.50 & -0.35 & - \\
& & & & & & \\
\hline
$\gamma p\rightarrow \psi p$ & -2.26 & 0.64 & - & 1.06 & 1.10 & - \\
& & & & & & \\
\hline
\end{tabular}
\vskip 1.cm

Actually, the high-energy part of $\sigma_{el}^{\gamma \rightarrow
J/\psi}$ is consistent with a moderate increase with $W$,
corresponding to a "soft" pomeron. The apparent rapid increase quoted
by different authors is a result of straight\-for\-ward in\-ter\-po\-lations 
from low to high energies without account for the nonasymptotic
contribution typical of $\sigma_{el}.$ In other words, the onset of
the logarithmic asymptotics occurs in $\sigma_{el}$ later than  in
$\sigma_{tot}$.  This is a general feature of the Regge pole or
geometrical models and it is shared by the above expression for
$\sigma_{el}$. As a consequence, in this type of models the ratio
$\sigma_{el}/\sigma_{tot}$ continues rising even after $\sigma_{tot}$
(but not $\sigma_{el}$) has reached its logarithmic asymptotics.

Notice also that the moderate, $ln(1/x)$ increase of the
photoproduction cross-sections is consistent with the recent data on
$F_{2}(x,Q^{2})$ from H1 \cite{H1-94}, and ZEUS that may be fitted by a
$ln(Q^{2})ln(1/x)$ type expression for not too small and large $Q^{2}$
\cite{Heidt}.

\medskip

\section{}
~

The presence of the diffractive (dip-bump) structure in the
differential cross section is an important indicative of 
diffraction. This structure is clearly seen in elastic hadronic
reactions, namely in $pp,\ \bar pp$ and $\pi p$ scattering.  The
details of this phenomenon, and in particular the evolution with energy
as well as the dependence on the masses (radii) of the colliding
particles have been studied in details \cite{J,BG}. The dip has been
predicted also in other elastic reactions.

Less clear is the dependence of the dip-bump structure on the
multiplicities (diffractive dissociation, DD) and virtualities (deep
inelastic scattering). Measurements of single DD at high
energies extend up to $t\sim -1 GeV^2$. No dip is visible within
this kinematical range. Little attention has been paid untill now to
the appearence of the dip in DD. We are sure that, because of the
univesal nature of diffracton the dip will show up sooner or later in
DD too.

Even less attention has been paid to the possibility of such a
structure in deep inelastic scattering. The first evidence of such an
event is of great intersest.

From most general arguments, the postion of the dip depends on two
factors: the slope of the first cone (the smaller the slope the
further the dip) and the strength of the rescattering (alternatively
-- the hight of the second cone). The slope in diffractive
$J/\psi$ production is about $4 GeV^{-2}$ - almost 1/4 of that in
$pp$, which has the effect of shifting the dip far away towards
large $|t|$. The possible existence of the dip at values of $|t|$ even
smaller than in $pp$ means strong absorption effects (unitarity
corrections) at large $\tilde q^2$, that may  compensate the above
trend and shift the dip back to small $|t|$.
\vskip 1.cm
\begin{minipage}[t]{8.cm}
\begin{center}
\epsfig{file=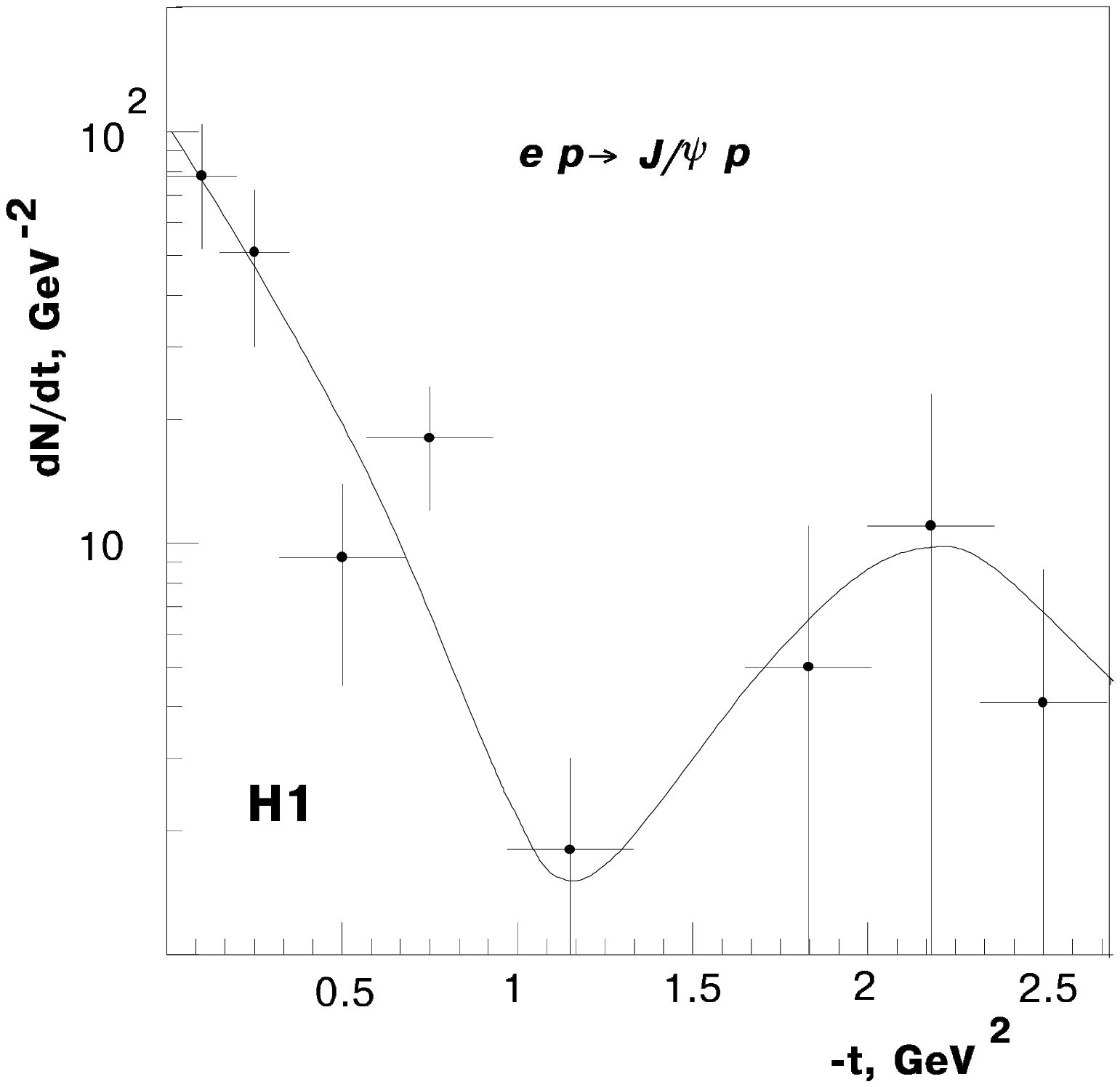,width=7.cm}
\vskip -.5cm
\end{center}
Fig.4. Possible dip-bumb structure in $J/\psi$-production.
\end{minipage} \ \
\begin{minipage}[t]{8.cm}
\vskip -8.cm
\hskip .5cm
In Fig.4 we present the
HERA \cite{dip} data on the $t-$dependence of the diffractive $J/\psi$
production. The curve is drown to guide the eye.
The studies of this interesting phenomenon open a new page in diffractive
studies.

\hskip .5cm
Another interesting class of re\-ac\-tions where the dip should also
appear is dif\-frac\-tive deep inelastic scattering, where the
mea\-su\-re\-ments have recently reached values of about $t=-1 GeV^2$.
\end{minipage}

\medskip \section{}
~

The main conclusion of the present paper is that the apparent rapid
increase of $\sigma^{\gamma\rightarrow J/\psi}_{el}(s)$  above
$\sqrt{s}\approx 20GeV$ is a preasymptotic effect. Asymptotically, this
cross-section will level off and will not exceed the rise of
$\sigma^{\gamma p}_{tot}$.

We thank M.Arneodo for usefull correspondence and valuabe discussions.


\medskip

\begin{thebibliography}{99}
\bibitem {Collins} P.D.B.Collins, An Introduction to Regge Theory \& High Energy
Physics, Cambridge Univ. Press, 1977.
\bibitem {JMP} L.Jenkovszky, E.Martynov and F.Paccanoni, {\it Regge behaviour
 of the nucleon structure function}, Padova preprint, DFPD 95/TH/21.
\bibitem {Cap} A.Capella, A.Kaidalov, C.Merino and Tran Tahn Van, Phys. Lett.
 {\bf B337}(1994)358.
\bibitem {Haak} L.P.A.Haakman, A.Kaidalov and J.H.Koch, Phys. Lett. {\bf B365}
(1996)411.
\bibitem {DL} A.Donnachie and P.V.Landshoff, Phys. Lett. {\bf B348}(1995)213.
\bibitem {Buch} W.Buchm\"uller and A.Hebbecker, Phys. Lett. {\bf 355}
(1995)573.
\bibitem {Amun} J.F.Amundson, Phys. Lett. {\bf B372}(1996)127. 
\bibitem {H1} S.Aid et al. (H1 Collaboration) Preprints DESY 95-251, 1995;
96-023, 1996.
\bibitem {ZEUS} M.Derrick et al. (ZEUS Collaboration) Preprints DESY 95-143,
1995; 96-002, 96-037, 1996.
\bibitem {H1-94} S.Aid et al. (H1 Collaboration) Preprint DESY 96-039, 1996.
\bibitem {Heidt} W.Buchm\"uller and D.Heidt, Preprint DESY 96-061, 1996.
\bibitem {J} L.Jenkovszky, Riv. Nuovo Cim. {\bf 10}(1987)1.
\bibitem {BG} M.M.Bertini and
M.Giffon, Fiz. Elem. Chast. At. Yadra {\bf 26}(1995)32.
\bibitem {dip} S.Aid et al. (H1 Collaboration) Preprint DESY 96-037,
1996.
\end{thebibliography}
\end{document}